\documentclass[aps,prl,twocolumn,showpacs,preprintnumbers,amsmath,superscriptaddress,amssymb,floats,nofootinbib]{revtex4}
\usepackage[usenames]{color}
\usepackage{graphicx}
\begin{document}

\def\Journal#1#2#3#4{{#1} {\bf #2}, #3 (#4)}
\def\NCA{\rm Nuovo Cimento}
\def\NPA{{\rm Nucl. Phys.} A}
\def\NIM{\rm Nucl. Instrum. Methods}
\def\NIMA{{\rm Nucl. Instrum. Methods} A}
\def\NPB{{\rm Nucl. Phys.} B}
\def\PLB{{\rm Phys. Lett.}  B}
\def\PRL{\rm Phys. Rev. Lett.}
\def\PRD{{\rm Phys. Rev.} D}
\def\PRC{{\rm Phys. Rev.} C}
\def\ZPC{{\rm Z. Phys.} C}
\def\JPG{{\rm J. Phys.} G}


\title{The extraction of $\phi-N$ total cross section from 
$d(\gamma,pK^{+}K^{-})n$}

\newcommand*{\DUKE}{Duke University, Durham, North Carolina 27708}
\newcommand*{\DUKEindex}{1}
\affiliation{\DUKE}
\newcommand*{\OHIOU}{Ohio University, Athens, Ohio  45701}
\newcommand*{\OHIOUindex}{2}
\affiliation{\OHIOU}
\newcommand*{\SACLAY}{CEA, Centre de Saclay, Irfu/Service de Physique Nucl\'eaire, 91191 Gif-sur-Yvette, France}
\newcommand*{\SACLAYindex}{3}
\affiliation{\SACLAY}
\newcommand*{\JLAB}{Thomas Jefferson National Accelerator Facility, Newport News, Virginia 23606}
\newcommand*{\JLABindex}{4}
\affiliation{\JLAB}
\newcommand*{\SCAROLINA}{University of South Carolina, Columbia, South Carolina 29208}
\newcommand*{\SCAROLINAindex}{5}
\affiliation{\SCAROLINA}
\newcommand*{\NOWMIT}{Massachusetts Institute of Technology,
  Cambridge, Massachusetts  02139-4307}
\newcommand*{\NOWMITindex}{6}
\affiliation{\NOWMIT}
\newcommand*{\ANL}{Argonne National Laboratory}
\newcommand*{\ANLindex}{7}
\affiliation{\ANL}
\newcommand*{\ASU}{Arizona State University, Tempe, Arizona 85287-1504}
\newcommand*{\ASUindex}{8}
\affiliation{\ASU}
\newcommand*{\UCLA}{University of California at Los Angeles, Los Angeles, California  90095-1547}
\newcommand*{\UCLAindex}{9}
\affiliation{\UCLA}
\newcommand*{\CSU}{California State University, Dominguez Hills, Carson, CA 90747}
\newcommand*{\CSUindex}{10}
\affiliation{\CSU}
\newcommand*{\CMU}{Carnegie Mellon University, Pittsburgh, Pennsylvania 15213}
\newcommand*{\CMUindex}{11}
\affiliation{\CMU}
\newcommand*{\CUA}{Catholic University of America, Washington, D.C. 20064}
\newcommand*{\CUAindex}{12}
\affiliation{\CUA}
\newcommand*{\CNU}{Christopher Newport University, Newport News, Virginia 23606}
\newcommand*{\CNUindex}{13}
\affiliation{\CNU}
\newcommand*{\UCONN}{University of Connecticut, Storrs, Connecticut 06269}
\newcommand*{\UCONNindex}{14}
\affiliation{\UCONN}
\newcommand*{\ECOSSEE}{Edinburgh University, Edinburgh EH9 3JZ, United Kingdom}
\newcommand*{\ECOSSEEindex}{15}
\affiliation{\ECOSSEE}
\newcommand*{\FU}{Fairfield University, Fairfield CT 06824}
\newcommand*{\FUindex}{16}
\affiliation{\FU}
\newcommand*{\FIU}{Florida International University, Miami, Florida 33199}
\newcommand*{\FIUindex}{17}
\affiliation{\FIU}
\newcommand*{\FSU}{Florida State University, Tallahassee, Florida 32306}
\newcommand*{\FSUindex}{18}
\affiliation{\FSU}
\newcommand*{\GWU}{The George Washington University, Washington, DC 20052}
\newcommand*{\GWUindex}{19}
\affiliation{\GWU}
\newcommand*{\ECOSSEG}{University of Glasgow, Glasgow G12 8QQ, United Kingdom}
\newcommand*{\ECOSSEGindex}{20}
\affiliation{\ECOSSEG}
\newcommand*{\ISU}{Idaho State University, Pocatello, Idaho 83209}
\newcommand*{\ISUindex}{21}
\affiliation{\ISU}
\newcommand*{\INFNFR}{INFN, Laboratori Nazionali di Frascati, 00044 Frascati, Italy}
\newcommand*{\INFNFRindex}{22}
\affiliation{\INFNFR}
\newcommand*{\INFNGE}{INFN, Sezione di Genova, 16146 Genova, Italy}
\newcommand*{\INFNGEindex}{23}
\affiliation{\INFNGE}
\newcommand*{\INFNRO}{INFN, Sezione di Roma Tor Vergata, 00133 Rome, Italy}
\newcommand*{\INFNROindex}{24}
\affiliation{\INFNRO}
\newcommand*{\ORSAY}{Institut de Physique Nucl\'eaire ORSAY, Orsay, France}
\newcommand*{\ORSAYindex}{25}
\affiliation{\ORSAY}
\newcommand*{\ITEP}{Institute of Theoretical and Experimental Physics, Moscow, 117259, Russia}
\newcommand*{\ITEPindex}{26}
\affiliation{\ITEP}
\newcommand*{\JMU}{James Madison University, Harrisonburg, Virginia 22807}
\newcommand*{\JMUindex}{27}
\affiliation{\JMU}
\newcommand*{\KYUNGPOOK}{Kyungpook National University, Daegu 702-701, Republic of Korea}
\newcommand*{\KYUNGPOOKindex}{28}
\affiliation{\KYUNGPOOK}
\newcommand*{\MOSCOS}{Moscow State University, Moscow, Russia}
\newcommand*{\MOSCOSindex}{29}
\affiliation{\MOSCOS} 
\newcommand*{\UNH}{University of New Hampshire, Durham, New Hampshire 03824-3568}
\newcommand*{\UNHindex}{30}
\affiliation{\UNH}
\newcommand*{\NSU}{Norfolk State University, Norfolk, Virginia 23504}
\newcommand*{\NSUindex}{31}
\affiliation{\NSU}
\newcommand*{\ODU}{Old Dominion University, Norfolk, Virginia 23529}
\newcommand*{\ODUindex}{32}
\affiliation{\ODU}
\newcommand*{\RPI}{Rensselaer Polytechnic Institute, Troy, New York 12180-3590}
\newcommand*{\RPIindex}{33}
\affiliation{\RPI}
\newcommand*{\URICH}{University of Richmond, Richmond, Virginia 23173}
\newcommand*{\URICHindex}{34}
\affiliation{\URICH}
\newcommand*{\ROMAII}{Universita' di Roma Tor Vergata, 00133 Rome Italy}
\newcommand*{\ROMAIIindex}{35}
\affiliation{\ROMAII}
\newcommand*{\MOSCOW}{Skobeltsyn Nuclear Physics Institute, Skobeltsyn Nuclear Physics Institute, 119899 Moscow, Russia}
\newcommand*{\MOSCOWindex}{36}
\affiliation{\MOSCOW}
\newcommand*{\UNIONC}{Union College, Schenectady, NY 12308}
\newcommand*{\UNIONCindex}{37}
\affiliation{\UNIONC}
\newcommand*{\UTFSM}{Universidad T\'{e}cnica Federico Santa Mar\'{i}a, Casilla 110-V Valpara\'{i}so, Chile}
\newcommand*{\UTFSMindex}{38}
\affiliation{\UTFSM}
\newcommand*{\VIRGINIA}{University of Virginia, Charlottesville, Virginia 22901}
\newcommand*{\VIRGINIAindex}{39}
\affiliation{\VIRGINIA}
\newcommand*{\WM}{College of William and Mary, Williamsburg, Virginia 23187-8795}
\newcommand*{\WMindex}{40}
\affiliation{\WM}
\newcommand*{\YEREVAN}{Yerevan Physics Institute, 375036 Yerevan, Armenia}
\newcommand*{\YEREVANindex}{41}
\affiliation{\YEREVAN}

\newcommand*{\NOWUK}{University of Kentucky, Lexington, KY 40506}
\newcommand*{\NOWJLAB}{Thomas Jefferson National Accelerator Facility,
  Newport News, VA 23606}
\newcommand*{\NOWLANL}{Los Alamos National Laborotory, New Mexico, NM 87545}
\newcommand*{\NOWGWU}{The George Washington University, Washington, DC 20052}
\newcommand*{\NOWECOSSEE}{Edinburgh University, Edinburgh EH9 3JZ, United Kingdom}
\newcommand*{\NOWWM}{College of William and Mary, Williamsburg, VA 23187}

\author {X.~Qian}
\affiliation{\DUKE}
\author {W.~Chen}
\affiliation{\DUKE}
\author {H.~Gao}
\affiliation{\DUKE}
\author {K.~Hicks}
\affiliation{\OHIOU}
\author {K.~Kramer}
\affiliation{\DUKE}
\author {J.M.~Laget} 
\affiliation{\SACLAY}
\affiliation{\JLAB}
\author {T.~Mibe}
\affiliation{\OHIOU}
\author {S.~Stepanyan}
\affiliation{\JLAB}
\author {D.J.~Tedeschi}
\affiliation{\SCAROLINA}
\author {W. Xu}
\affiliation {\NOWMIT}
% ...

\collaboration{The CLAS Collaboration}
\noaffiliation

\date{\today}
\begin{abstract}
We report on the first measurement of the differential cross section of
$\phi$-meson photoproduction for the $d(\gamma,pK^{+}K^{-})n$ 
exclusive reaction channel. The experiment was performed 
using a \textcolor{black}{tagged-photon} beam and the
CEBAF Large Acceptance Spectrometer (CLAS) at Jefferson Lab. 
A combined analysis using data from the $d(\gamma,pK^{+}K^{-})n$ channel 
and those from a previous publication on 
coherent $\phi$ production on the deuteron
has been carried out to extract the $\phi-N$ total
cross section, $\sigma_{\phi N}$. The extracted $\phi-N$ total
cross section  
favors a value above 20 mb. This value is larger than 
the value extracted using vector-meson dominance models for $\phi$
photoproduction on the proton. 

\end{abstract}
\maketitle

Multi-gluon exchange between hadrons, known as Pomeron exchange, is a
fundamental process and plays an important role in high-energy
interactions. At lower energies, \textcolor{black}{this exchange}
manifests itself \textcolor{black}{in a QCD} van der Waals interaction~\cite{Brodsky:1990jd}. Studying
multi-gluon exchange at 
lower energies is challenging because at low energies hadron-hadron
interactions are dominated by quark exchange. 
However, multi-gluon exchange is expected
to be dominant in the interaction between two hadrons when
they have no common quarks. The $\phi$ meson is unique in that it is nearly
\textcolor{black}{an} $s\bar s$ state and hence gluon exchange is expected to dominate the 
$\phi -N$ scattering process.

Direct measurement of \textcolor{black}{the} $\phi-N$ cross section is not
  possible due to lack of $\phi$ meson beams.
An upper limit of $\sigma_{\phi N}\simeq 11$ mb~\cite{sibirtsev} is obtained using the
$\phi$ photoproduction data on the proton and the vector meson dominance
(VMD) model~\cite{Sakurai}, which is in agreement with the estimate from the additive
quark model~\cite{Lipkin}. In a geometric
  interpretation of hadron-proton total cross sections~\cite{povh},
  the radius of the $\phi$ meson $r_{\phi}$ can be estimated from the
  comparison of the total cross sections of $\pi-N$ ($\sigma_{\pi N}$)
  and $\phi-N$ ($\sigma_{{\phi}N}$) scattering. The value of 
  $\sigma_{\pi N}$ is $\sim 24$ mb~\cite{povh} and the $\pi$ radius $r_\pi$ is
  $\sim$ 0.65 fm~\cite{povh}. An upper limit of $\sigma_{{\phi}N} \sim
  11$ mb~\cite{sibirtsev} leads to an upper limit value of $\sim$ 0.43 fm for $r_\phi$.

However, from the observed $A$-dependence of nuclear $\phi$ photoproduction, a larger
value of (inelastic $\sigma^{inelas}_{\phi N}\simeq 35$
mb~\cite{leps}, which is part of the total $\sigma_{\phi N}$) is obtained, which suggests a
larger $r_\phi$ value than 0.43 fm. Medium modification of the vector meson
properties~\cite{medium} (such as radius) or channel coupling effects~\cite{sibirtsev}
have been proposed to explain the aforementioned difference in the
cross section for the $\phi$ meson. A similar phenomenon 
has been observed for the $J/\Psi$ meson~\cite{Mark1} and the color transparency
effect is proposed~\cite{Mark2} to explain the observation.

In this paper, we present \textcolor{black}{a} determination of $\sigma_{\phi N}$ using the
differential cross section of the incoherent $\phi$-meson photoproduction
from deuterium. This process takes advantage of the rescattering of a $\phi$ meson
from the spectator nucleon. In the  reaction $\gamma + d \to \phi + p + n$,
the rescattering process will dominate for the kinematics where both nucleons
are energetic. \textcolor{black}{The deuteron} is a system of loosely bound nucleons and, hence,
nuclear medium effects should not play a significant role in the $\phi -N$
scattering process. In the analysis of incoherent
  $\phi$-meson photoproduction from deuterium, the $\phi -N$
  interaction is parametrized as $\frac{d\sigma}{dt} \propto \sigma_{\phi
  N}^2 \cdot e^{\beta_{\phi N} t}$~\cite{Laget,Misak1,Misak2}, \textcolor{black}{where}
$\beta_{\phi N}$ characterizes the $t$-dependence of the differential cross
section, and $t$ is defined as the four-momentum transfer squared
between the photon and the $\phi$ meson. A $\chi^2$ analysis is
performed for both  
processes \textcolor{black}{($\gamma + d \to \phi + d$ as in \cite{mibe}
  and $\gamma + d \to \phi + p + n$) in this paper} to constrain the
values of $\sigma_{\phi N}$ and $\beta_{\phi N}$.  

The rescattering of a $\phi$ meson off a nucleon in the deuteron was
used to study the $\phi -N$ interaction in a recent analysis of CLAS g10 data
using coherent $\phi$ photoproduction, $\gamma + d \to \phi +
d$~\cite{mibe}. In the coherent $\phi$ production process,
rescattering dominates in the high $|t|$
region. \textcolor{black}{Results of~Ref.\cite{mibe} 
agree with the VMD prediction (where $\beta_{\phi N}$ is assumed to be the one
in the $\phi$-meson photoproduction from a nucleon),
however, larger $\sigma_{\phi N}$ values showed better agreement with
the data if one allowed the slope ($\beta_{\phi N}$)
of the $t$-dependence of the $\phi -N$ scattering process to be different.}

The reaction $\gamma(d,\phi p)n$ was measured 
by detecting kaons from the $\phi$-meson decay ($\phi \rightarrow
K^+ K^-$, branching ratio about 0.5) using 
the same data \textcolor{black}{set as in Ref.~\cite{mibe}.} A
tagged-photon beam was generated by a 3.8-GeV electron beam incident 
on a gold radiator ($10^{-4}$ radiation length). The
photon flux was measured by the CLAS photon-tagger system~\cite{sober}.
The data were collected from a 24-cm-long 
liquid-deuterium target in the CLAS
detector~\cite{CNIM} at JLab. 
 

Events having the final state $\gamma + d \rightarrow K^+ + K^- + p + X$
were selected using a triple coincidence detection of a proton, a
$K^{+}$ and, a $K^{-}$.  Each particle was selected based on 
particle charge, momentum, and Time-of-Flight (TOF) 
information. The reaction $d(\gamma,pK^+K^-)n$ was
identified in the missing mass squared distribution by the missing
neutron shown in Fig. 1a. In the figure, the position of the neutron
mass squared is shown by the dotted line. A 
$\pm3\sigma$ cut was
employed to select the $pK^+K^-(n)$ final state events. 


\begin{figure} 
\begin{center}
\includegraphics[width=90mm]{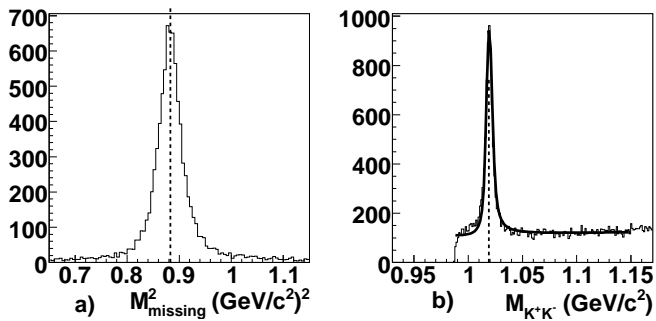}
\caption{(a) The missing mass squared distribution; (b)
  the invariant mass distribution for the $K^+ K^-$ for photon energies
  between 1.65 to 3.59 GeV for the $\gamma + d \rightarrow K^+ + K^- + p +
  X$ process. Also shown in (b) is a fit using a Breit-Wigner function 
  convoluted with the experimental resolution together with a 
  flat background.}  
\label{fig:com}
\end{center} 
\end{figure}

\textcolor{black}{Once the reaction $d(\gamma,pK^+K^-)n$ was identified, the number of
$\phi$ mesons was obtained by subtracting the background under the
$\phi$ peak (invariant mass spectrum of the $K^+$ and $K^-$)
in the $\pm 3\sigma$ region (see Fig. 1b).} 
The $K^+K^-$ invariant mass distribution 
was fitted using a Breit-Wigner function convoluted with the 
experimental resolution, plus a function to model the background in each
photon energy and $|t-t_0|$ bin, where $t_{0}$ is the minimum
$t$ value for a given photon energy. The background shape was assumed
to be~\cite{mibe}: 
\begin{eqnarray}
f(x) & = & a \sqrt{x^{2} - (2M_{K})^{2}} + b\left (x^{2}-(2M_{K})^{2}
\right ) (x > 2M_{K}) \nonumber\\
f(x) & = & 0~(x<2M_{K}),
\end{eqnarray}
where $x$ is the invariant mass of the $K^{+}K^{-}$, $M_K$ is the kaon mass, 
and $a$ and $b$ are fitting parameters. \textcolor{black}{ The
  background was also fitted with a linear shape. The results from
  fitting these two shapes were compared to estimate the systematic
  uncertainties due to the subtraction of the background.} 


\textcolor{black}{A Monte Carlo (MC)~\cite{gsim} simulation of the CLAS
  detector was carried out to determine the efficiency for the
  detection of the $\gamma + d \to p + K^+ + K^-$ reaction.} Two event generators
were used in two different missing neutron momentum regions. 
A quasi-free event generator for the $\gamma + p \rightarrow \phi + p$
process in the deuteron was employed for \textcolor{black}{the missing
  neutron momentum} distribution below 0.18 GeV/c, where the
  agreement between the MC simulation and the data is very good. 
The deuteron wave function was based on the Bonn
potential~\cite{bonn}.  The differential cross sections from
CLAS~\cite{g11dxs} for $\phi$ photoproduction from the proton were used.  
For \textcolor{black}{the missing neutron momentum} greater than $0.18$
GeV/c, the generated events 
were weighted by $\frac{d\sigma}{dP_nd\Omega_nd\Omega_p}$ 
from ~Ref.\cite{Laget} to  
include both the $\phi-N$ and $N-N$ final state interactions.   
In both cases, the $\phi$ mass distribution was modeled by a Breit-Wigner shape with the 
resonance centered at the $\phi$ mass of 1.019 GeV/c$^{2}$ and 
with a FWHM of $0.0044~$GeV/c$^{2}$. 
The $\phi$-meson decay angular distribution was taken as~\cite{angular1}:
\begin{equation}
W(\cos(\theta_H))=\frac{3}{2}(\frac{1}{2}(1-\rho^0_{00})\sin^2\theta_H +
\rho^0_{00}\cos^2\theta_H) + \alpha \cos\theta_H ,
\end{equation}
where $\rho^0_{00}$ is the spin density matrix element, 
$\theta_H$ denotes the polar
angle of the $K^+$ in the $\phi$-meson rest frame, and $\alpha$
accounts for an interference between the $\phi$ and the 
non-resonant
$S$-wave $K^+K^-$ pair production~\cite{prephi}. Helicity
conserving amplitudes give $\rho^0_{00} = 0$, while single-helicity
flip amplitudes require $\rho^0_{00} \neq 0$. A value of 0.1 (0.05) was used
for $\rho^0_{00}$ ($\alpha$) in the MC simulation.
The MC generated events were used as input to the GEANT3-based CLAS
simulation~\cite{gsim}. 
They were then reconstructed using the same event reconstruction 
algorithm as was used for the data.
The acceptance was obtained by the ratio of \textcolor{black}{ the number of events 
that passed the analysis cuts to the number of 
generated $\phi$ events}. The average differential cross section for
each photon energy and $|t-t_0|$ bin was
extracted by dividing the normalized yield (number of selected events
divided by the integrated photon flux including the DAQ dead time, the
target thickness, the $\phi$ decay branching ratio
and $|t-t_0|$ bin size) by the acceptance which includes the detection
efficiency. The differential cross sections were then bin-centered at
fixed $t$ values and a finite binning correction was applied.

Several sources contribute to the overall systematic uncertainty in
extracting the differential cross section.
The systematic uncertainties associated with particle identification 
and the missing mass cut were determined to be 0.5 - 7.0\% and 0.5 -
5.0\%, \textcolor{black}{which are the values found across the different
bins of photon energy and $t$}, respectively. These were
determined by varying the corresponding  
cuts by $\pm 10$\%. The
uncertainties in the parameters of the $\phi$ decay angular
distribution, $\rho^0_{00}$ and $\alpha$, 
were 10\% and 5\%~\cite{g11dxs,prephi}, respectively,
leading to 1.5-6.5\% systematic uncertainties. The background obtained
from the non-linear background shape was on average 5\% smaller than
that from  the linear background. The systematic
uncertainties from the acceptance dependence on the cross section
model varied from 0.5\% to 9\%. The uncertainty in the photon flux was
5\%~\cite{gflux,weinote}. The uncertainty of the bin-centering corrections were
typically between 0.5\% and 6.0\%, based on our current knowledge of
the CLAS detector and the uncertainty in the input cross section. 
Combining all systematic uncertainties in quadrature, 
the overall systematic uncertainties vary from $7\% - 18\%$ depending on the 
kinematics.

\begin{figure} 
\begin{center}
\includegraphics[width=90mm]{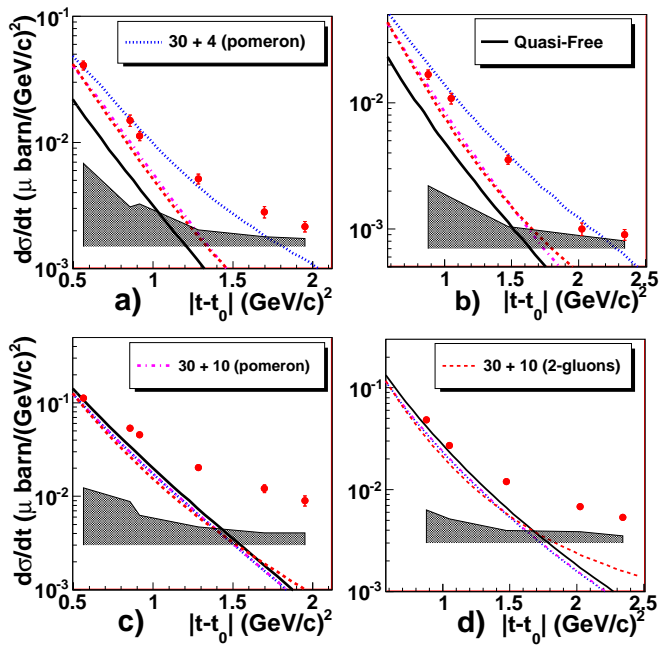}
\caption{Differential cross section ${\frac{d\sigma}{dt}}$ from
  $\gamma(d,K^+K^-p)n$ process \textcolor{black}{for photon energies of
  1.65-2.62 GeV (a, c) and 2.62-3.59 GeV (b, d). The missing momentum
  is higher than 0.18 GeV/c in (a) and (b), and lower than 0.18 GeV/c
  in (c) and (d).} 
  \textcolor{black}{The results of this work are shown in red solid
  circles. The black bands represent the systematic uncertainties.} 
  The label ``30 + 10'' indicates the calculation from 
  Laget~\cite{Laget} with $\sigma_{tot}^{\phi N}$ = 30 mb and
  $\beta_{\phi N}$ = 10 (GeV/c)$^{-2}$. The legend for the
  calculations is presented for better visibility and it applies to
  all four panels. } 
\label{fig:result}
\end{center} 
\end{figure}

In  Fig.~\ref{fig:result}, differential cross sections,
$\frac{d\sigma}{dt}$ \textcolor{black}{from this work (red solid
  circles) for the reaction } $\gamma + d \rightarrow \phi + p + n$
are presented. 
In Fig. 2a and Fig. 2b, we present ${\frac{d\sigma}{dt}}$ as a function of 
 $|t-t_0|$ for a photon energy range of 1.65 - 2.62 GeV and 2.62 - 3.59 GeV,
respectively (same as those in ~Ref.\cite{mibe}).  
\textcolor{black}{The detected proton and the missing
  neutron momentum span a range} of 0.18 to 2.0 GeV/c (Figs.~2a and ~2b).
For the low missing momentum region
(the momenta of the reconstructed neutron smaller than 0.18 GeV/c),
cross sections over the same photon energy ranges 
are shown in Fig. 2c and Fig. 2d.   
\textcolor{black}{The error bars shown are the statistical uncertainties, 
and the overall systematic uncertainties are shown by the black band.  
Also plotted are the
predictions~\cite{Laget} for quasi-free $\phi$ production and
rescattering for four sets of $\sigma_{\phi N}$ and $\beta_{\phi
  N}$. The calculations are performed based on the model assumptions of
pomeron exchange or the two-gluon exchange interaction in the $\phi-N$
rescattering process.}

In~\cite{Laget}, the
neutron-proton rescattering amplitude has been taken into account and
treated in the same way as in the analysis of the 
 $d(e,e'p)n$ channel~\cite{Laget1,Egiyan}.
The models for the $\phi$ photoproduction on the
proton~\cite{Laget2,Cano} used in Ref.~\cite{Laget} describe  
the published experimental cross sections~\cite{prephi,Anciant} reasonably
well for photon energies in the vicinity of 3.4 GeV and above.
However, at low photon energies the model underestimates the
experimental cross sections of $\gamma + p \rightarrow \phi +
p$~\cite{g11dxs} above $|t-t_0|$ = 1 (GeV/c)$^2$ 
by a factor that can reach 10 at 2.5 (GeV/c)$^2$.
 The extracted
cross sections~\cite{qian2} of the $\gamma + ``p'' \rightarrow p + \phi$ from 
the $d(\gamma,\phi p)n$ channel at low neutron missing momentum 
from this work are in
good agreement with the extracted cross sections of $\gamma + p
\rightarrow \phi + p$~\cite{g11dxs}.
The flattening behavior of the experimental cross sections with increasing 
$|t-t_0|$ values
is well accounted for by the coupling between the $\phi$ and $\omega$ 
production channels~\cite{laget-mixing}, which is discussed later. This channel
coupling effect has not yet been 
implemented in Laget's code that we use, 
and its effect will be investigated in a future study. 

We simply note that in the high missing momentum region where contributions of
rescattering processes ($\phi -N$ or $N-N$) are significant, 
the dominant contribution to the matrix element comes from the
photoproduction of $\phi$ on a nucleon at rest, which gets its measured
momentum in the rescattering
process (we refer to Ref.~\cite{Laget} for a detailed discussion). The consequence is
that in the scattering loop the
elementary photoproduction amplitude $\gamma + N\to \phi + N$ is 
almost the same
as in the quasifree case, when the spectator is at rest. 
The amplitude for the $\gamma + d \rightarrow \phi + p +
n$ process can be written as: 
\begin{widetext}
\begin{equation}
A = A_{\phi p}^{QS} \otimes (1 + A_{pn}^{fsi} +
A_{\phi n}^{fsi}) + A_{\phi n}^{QS} \otimes (1 + A_{pn}^{fsi} +
A_{\phi p}^{fsi}) 
 = (A_{\phi p}^{QS} + A_{\phi n}^{QS}) \otimes (1 + A_{pn}^{fsi} +
A_{\phi N}^{fsi})
\end{equation}
\end{widetext}
where $A_{\phi N}^{QS}$, $A_{pn}^{fsi}$ and $A_{\phi N}^{fsi}$ are
the amplitude for the $\phi$-meson photoproduction from nucleon, 
the amplitude for the 
proton-neutron final state interaction, and the amplitude for the
$\phi$-N final state interaction, respectively. 
The second step assumes isospin
symmetry for the $\phi$-N interaction. The $\otimes$ represents a
convoluted integral over internal momentum.  Thus, deviations from data 
in the model cross section of $\phi$ photoproduction from nucleons will lead to
deviations of the calculated cross sections in the high spectator nucleon 
momentum region from the data, even without 
final state interaction effects. In order to
minimize this effect, we form the following ratio between the results
with a high spectator momentum cut and the results with a low spectator
momentum cut for each value of $|t-t_0|$: 
\begin{equation} 
R = \frac{\sigma_{H}}{\sigma_{L}} = \frac{\int^{2 GeV}_{0.18
GeV}\frac{d\sigma}{dtdP_{miss}}\cdot dP_{miss}}{\int^{0.18 GeV}_{0.
GeV}\frac{d\sigma}{dtdP_{miss}}\cdot dP_{miss}}
\end{equation}
The amplitude with a low spectator nucleon momentum cut is 
\begin{equation}
A = A_{\phi p}^{QS} + A_{\phi n}^{QS}
\end{equation}
within the quasi-free reaction mechanism.
Therefore, the extracted ratio is 
\begin{eqnarray}
R &\sim& \frac{|(A_{\phi p}^{QS} + A_{\phi n}^{QS}) \otimes (1 + A_{pn}^{fsi} +
A_{\phi N}^{fsi})|^2}{|(A_{\phi p}^{QS} + A_{\phi n}^{QS})|^2}
\\\nonumber
&\sim& |(1 + A_{pn}^{fsi} + A_{\phi N}^{fsi})|^2
\end{eqnarray}
The second step relies on the factorization approximation which works
better when the elementary amplitude varys slowly with energy. This is
the case for the $\phi$-meson photoproduction channel. 
\textcolor{black}{This implies that to first order in the ratio $R$, 
the elementary $\phi$
photoproduction amplitude and, hence, the related model uncertainties, should
cancel out.} 


The comparison of the experimental ratios to the model calculations
from~Ref.\cite{Laget} are presented in Fig. 3. The experimental data are shown with
statistical uncertainties. The systematic uncertainties are 
shown by the black band.
In the low energy range,
$E_\gamma$ = 1.65-2.62 GeV (Fig.~3a), the data are best described by the parameters of 
$\sigma_{\phi N} = 10$ mb and $\beta_{\phi N}=6$
(GeV/c)$^{-2}$. In the range $E_\gamma$ = 2.62-3.59 GeV, the data can
be described well by all four calculations shown in Fig. 3 including
the rescattering effect.

\begin{figure*} 
\begin{center}
\resizebox{5in}{2.56in}{\includegraphics{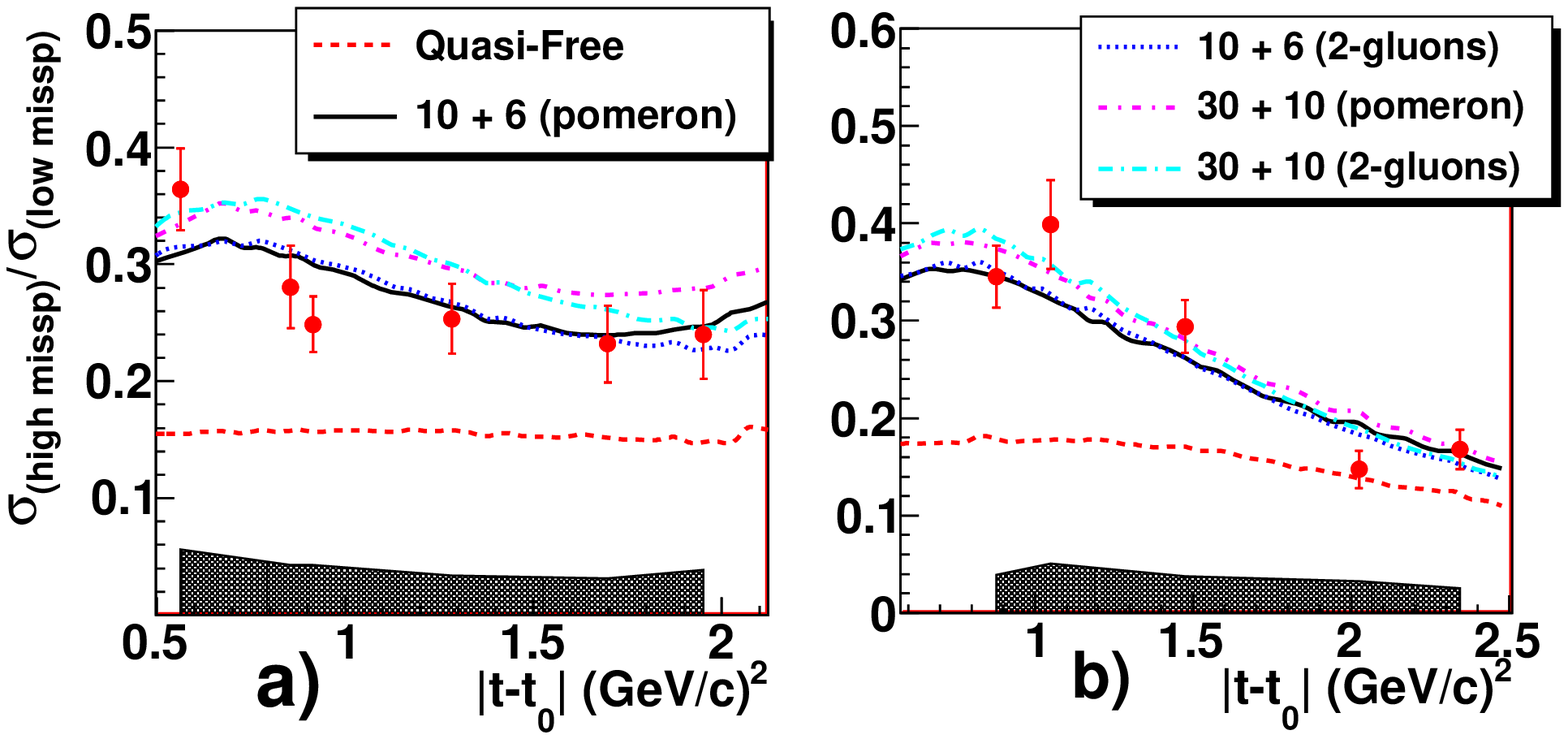}}
\caption{Cross section ratio between the high and the low missing momentum
  regions for photon energies of 1.65-2.62 GeV (a) and 2.62-3.59 GeV
  (b). \textcolor{black}{The results of this work are shown in red solid circles. The black bands
  represent the systematic uncertainties. We use same notations as
  those in Fig.~\ref{fig:result} for calculations from Laget~\cite{Laget}.}}
\label{fig:result1}
\end{center} 
\end{figure*}


In order to constrain the value of $\sigma_{\phi N}$ using our data, a $\chi^2$ analysis
was performed for results from both the $\gamma + d \rightarrow \phi + p +
(n)$ process ($R = \frac{\sigma_{H}}{\sigma_{L}}$) and the published  
$\gamma + d \rightarrow \phi + d$ coherent channel~\cite{mibe} by
mapping the phase space of $\sigma_{\phi N}$ and $\beta_{\phi
  N}$. The $\chi^2$ is defined as:
\begin{equation}
\chi^2 = \sum_{i=1}^{N} \frac{(R_{data} - R_{cal})^2}{\delta R_{data}^2},
\end{equation}
where $N$ is the number of data points. $R_{data}$ ($R_{cal}$) is the
cross section ratio defined in Eq.~3  
for data (calculation). 
The $\delta R_{data}$ is quadrature sum of the point-to-point systematic uncertainty and the
statistical uncertainty. 
For the coherent process, $R_{data}$ ($R_{cal}$)
is the differential cross section data (calculation). 
The calculations of Laget~\cite{Laget}, which include 
pomeron exchange for the elementary $\phi$ meson photoproduction
cross section on the nucleon, are used for the $\gamma + d \rightarrow
\phi + p +n$ channel. The pomeron and the two-gluon exchange
versions of the model lead to very similar results at $|t-t_0|$ values
smaller than 1.5 (GeV/c)$^2$, as can be seen in Fig.~2 and Fig.~3. 
For the coherent production channel, calculations from Sargsian {\it
  et al.}~\cite{Misak1,Misak2} are used.


%

 
Fig. 4a and Fig. 4b show the confidence level for both processes in the two
photon energy regions. 
While the energy dependence in $\sigma_{\phi N}$ and $\beta_{\phi N}$ might be a 
possible explanation for the difference  between Fig.~4a and 4b,
the combined analysis favors a value of $\sigma_{\phi N}$ larger
than 20 mb. Our results are consistent with that extracted from the
SPring-8 data~\cite{leps} in which Li, C, Al and Cu nuclear targets
were used. Further more, our combined analysis gives a lower bound of 6
(GeV/c)$^{-2}$ for the $\beta_{\phi N}$ parameter. 

\begin{figure*} 
\begin{center}
\resizebox{4.5in}{2.3in}{\includegraphics{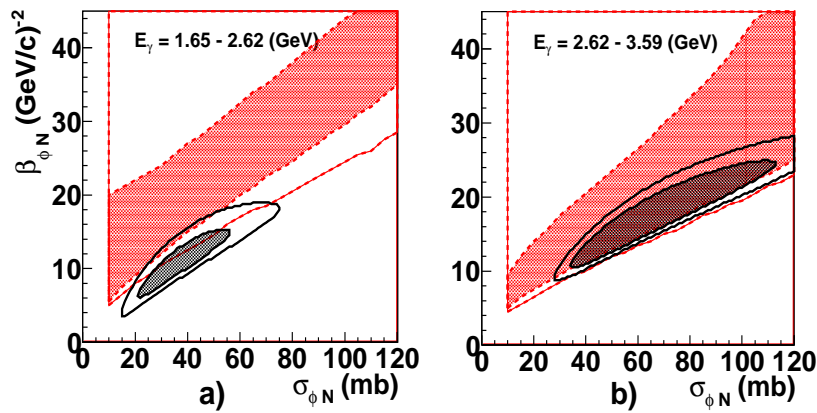}}
\caption{ \textcolor{black}{The 70\% (shaded area) and the 95\% (open
  area) confidence level plots shown   
  for the $\gamma + d \rightarrow \phi + p + (n)$ channel (red), the 
  $\gamma + d \rightarrow \phi + d$ coherent channel (black) for
  photon energies of 1.65-2.62 GeV (a) and 2.62-3.59 GeV (b).}
%
}
\label{fig:conf}
\end{center} 
\end{figure*}

Medium modifications that have been suggested by the SPring-8
data~\cite{leps} can hardly explain a large $\sigma_{\phi N}$ cross
section (large $r_\phi$) in deuterium. More likely it may reflect the fact that other
mechanisms, beyond $np$ and $\phi N$ rescattering, are at play and are
more important than the medium modifications, for example the QCD van
der Waals force. On the one hand, the coupling of the $\phi$ (via
two-gluon exchange) to hidden color components~\cite{lagethiddencolor}
inside the deuteron may contribute to large missing momenta and leave
less room for $\phi-N$ rescattering. On the other hand, the coupling to 
a cryptoexotic baryon (baryon with hidden strangeness),
$B_\phi=udds\bar{s}$, may also contribute~\cite{sibirtsev}. \textcolor{black}{However,} the
most likely explanation lies in the $\omega-\phi$ mixing. The photon produces an
$\omega$ meson on one nucleon, which elastically scatters on the second nucleon
before transforming into a $\phi$ meson. The corresponding matrix element has
the same structure as the elastic $\phi$ rescattering matrix element that we
considered in our analysis. 
An effective $\sigma_{\phi N}$ cross section value can be written as:
\begin{equation}
\sigma^{eff}_{\phi N}\sim \sigma_{\phi N} + \sigma_{\omega N}
\sqrt{\frac{\sigma_{\gamma N \rightarrow \omega N}}{\sigma_{\gamma N \rightarrow \phi N}}}
g_{\omega\rightarrow\phi}.
\end{equation}
With the experimental value  of 
${\frac{\sigma_{\gamma N \rightarrow \omega N}}{\sigma_{\gamma N \rightarrow 
\phi N}}} \sim 50$~\cite{Anciant,omega} 
in a $-t$ region of 1 to 3 
(GeV/c)$^2$, and the $\omega$-$\phi$ mixing coefficient $g_{\omega
  \rightarrow \phi} \sim 0.09$~\cite{pdg}, 
one can reconcile the effective $\phi N$ cross section of $\sim$ 30
mb with the VMD values of the $\sigma_{\omega N}$ ($\sim$ 25 mb) and 
$\sigma_{\phi N}$ ($\sim$ 10 mb)~\cite{laget-mixing}.
One way to put the $\phi-N$ cross section on more solid ground would be to
select the part of the phase space where the $\phi-N$ rescattering
dominates, using the method proposed in
Ref.~\cite{Laget}. Future high statistics data from a luminosity upgraded 
CLAS12~\cite{clas12} detector will help disentangle these possibilities.
Furthermore, future improved and new theoretical calculations will allow us 
to study the model uncertainty in the extraction of the $\phi$-N total cross section.






\textcolor{black}{
We thank Misak Sargsian and Mark Strikman for
  helpful conversations. We acknowledge the outstanding efforts of the staff
of the Accelerator and Physics
Divisions at Jefferson Lab who made this experiment possible.
This work was supported in part by the U.S.~Department of Energy under
contract number DE-FG02-03ER41231, the National Science Foundation,
the Istituto Nazionale di Fisica Nucleare,  
  the French Centre National de la Recherche Scientifique, 
  the French Commissariat \`{a} l'Energie Atomique, 
  the U.S. Department of Energy,  
  the National Science Foundation,  
  the UK Science and Technology Facilities Council (STFC),
  and the Korean Science and Engineering Foundation.  The Southeastern
  Universities Research Association (SURA) operates the  
 Thomas Jefferson National Accelerator Facility for the United States 
 Department of Energy under contract DE-AC05-84ER40150.
}

\end{document}